\documentclass[12pt]{iopart}%one column

\usepackage{iopams}
\usepackage{graphicx}% Needed to include png/jpg/pdf figure files
\usepackage{bm}
\usepackage{color}

\begin{document}

\title{Dark excitons and the elusive valley polarization in transition metal dichalcogenides}

\author{M Baranowski $^{1,2}$\footnotemark[1], A Surrente$^1$\footnotemark[1], D K\ Maude$^1$, M Ballottin$^3$, A A Mitioglu$^3$, P C M Christianen$^3$, Y C Kung$^4$, D Dumcenco$^4$, A Kis$^4$ and P Plochocka$^{1}$ }

\address{$^1$Laboratoire National des Champs Magn\'etiques Intenses, UPR 3228, CNRS-UGA-UPS-INSA, Grenoble and Toulouse, France}
\address{$^2$ Department of Experimental Physics, Faculty of Fundamental Problems of Technology, Wroclaw University of Science and
Technology, Wroclaw, Poland}
\address{$^3$High Field Magnet Laboratory (HFML -- EMFL), Radboud University, 6525 ED Nijmegen, The Netherlands}
\address{$^4$Electrical Engineering Institute and Interdisciplinary Center for Electron Microscopy (CIME), \'Ecole Polytechnique
F\'ed\'erale de Lausanne (EPFL), CH-1015 Lausanne, Switzerland}
\ead{paulina.plochocka@lncmi.cnrs.fr}
\vspace{10pt}
\begin{indented}
\item[]November 2016
\end{indented}

\footnotetext{these authors contributed equally to this work}

\begin{abstract}
A rate equation model for the dark and bright excitons kinetics is proposed which explains the wide variation in the observed
degree of circular polarization of the PL emission in different TMDs monolayers. Our work suggests that the dark exciton states
play an important, and previously unsuspected role in determining the degree of polarization of the PL emission. A dark exciton
ground state provides a robust reservoir for valley polarization, which tries to maintain a Boltzmann distribution of the bright
exciton states in the same valley via the intra valley bright dark exciton scattering mechanism. The dependence of the degree of
circular polarization on the detuning energy of the excitation in MoSe$_2$ suggests that the electron-hole exchange interaction
dominates over two LA phonon emission mechanism for inter valley scattering in TMDs.
\end{abstract}

% Uncomment for PACS numbers
%\pacs{00.00, 20.00, 42.10}
%
% Uncomment for keywords
%\vspace{2pc}
%\noindent{\it Keywords}: XXXXXX, YYYYYYYY, ZZZZZZZZZ
%
% Uncomment for Submitted to journal title message
%\submitto{\JPA}
%
% Uncomment if a separate title page is required

\submitto{\TDM} \maketitle
%
% For two-column output uncomment the next line and choose [10pt] rather than [12pt] in the \documentclass declaration
%\ioptwocol%has to be after \maketitle
%

\section{Introduction}
Monolayer transition metal dichalcogenides (TMDs) have attracted tremendous interest due to their unique electrical and optical
properties that are absent in their bulk and fewlayer forms\,\cite{wang2012electronics, xu2014spin, rev2, rev1, rev5}. Bulk TMDs
such as MoS$_2$, MoSe$_2$, WS$_2$ and WSe$_2$ are indirect band gap semiconductors with the conduction band minimum located half
way between the $\mathbf{K}$ and $\mathbf{\Gamma}$ points of the Brillouin zone, while the valence band maximum is located at the
$\mathbf{\Gamma}$ point. Monolayer TMDs have a direct band gap with the conduction and valence band minima located at the the
$\mathbf{K}$ point of the Brillouin zone\,\cite{PhysRevLett.105.136805, zeng2013optical, splendiani2010emerging, rev5}. The
direct gap in the visible light energy range makes single-layer TMDs very attractive materials for optoelectronic devices such
as light emitting diodes and detectors\,\cite{lopez2014light, jariwala2014emerging, ross2014electrically, wang2012electronics,
zhang2014electrically, mak2016photonics}. Moreover, monolayer TMDs possess a unique band structure that makes them ideal material
systems for valleytronics\,\cite{xiao2010berry,xiao2012coupled,xu2014spin,cao2012valley}.

\begin{figure}[h!]
\centering
\includegraphics[width=0.8\linewidth]{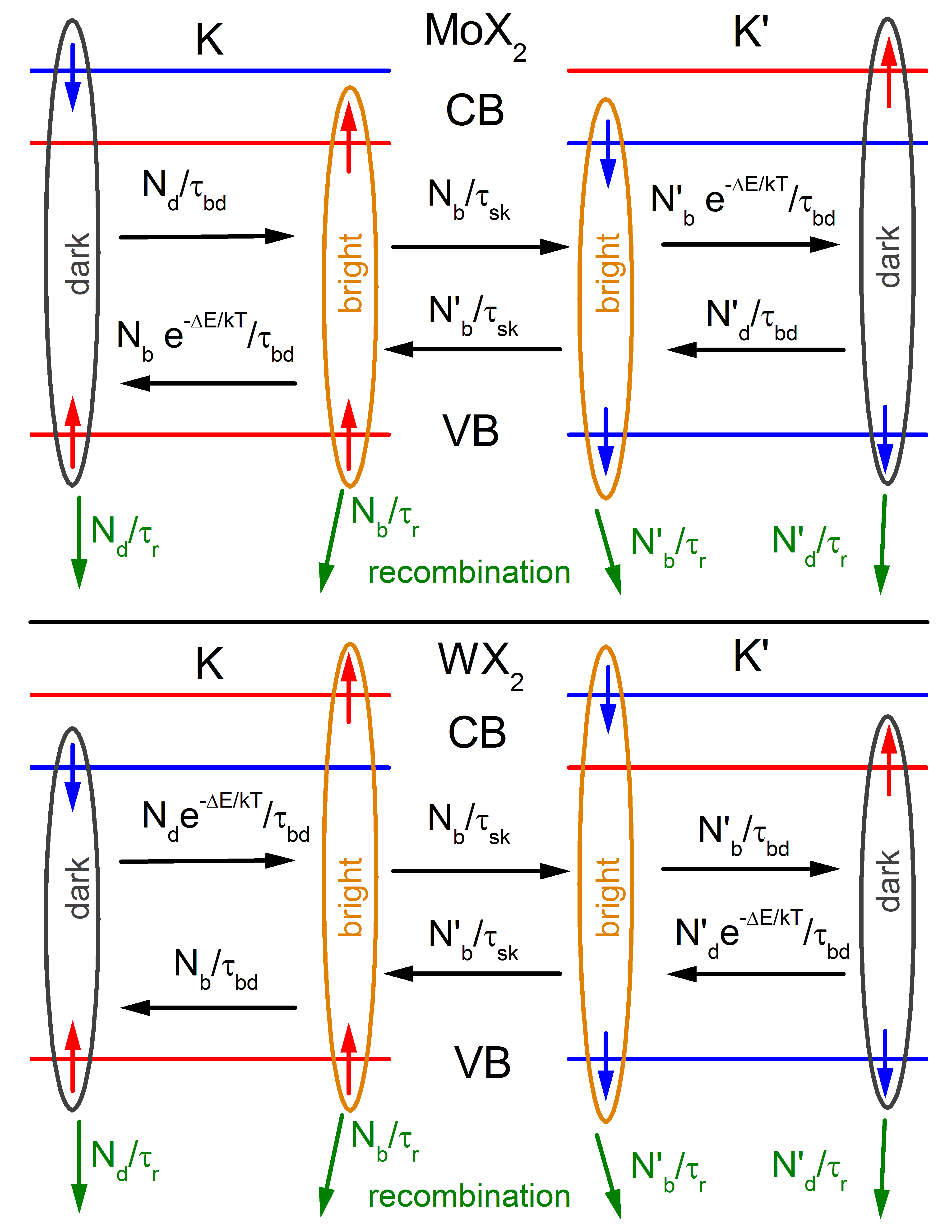}
\caption{Schematic showing the band structure and spin configuration of the bright and dark A-excitons in MoX$_2$ and WX$_2$
materials. The possible inter and intra valley scattering and recombination paths included in the rate equations (see text) are
indicated schematically. The influence of B-excitons can safely be neglected due to the large spin-orbit splitting in the valence
band.} \label{fig:scheme}
\end{figure}

The band-edges in monolayer TMDs are located at the degenerate but non equivalent $\mathbf{K}$ and $\mathbf{K'}$ corners of the
hexagonal Brillouin zone.  The valence band electrons in $\mathbf{K}$  and  $\mathbf{K'}$ valleys have non zero magnetic moment
related to hybridization of the $d_{x^2-y^2}$ and $d_{xy}$ orbitals\,\cite{liu2013three}. Broken inversion symmetry together with
time reversal symmetry lead to magnetic moments in the $\mathbf{K}$ and $\mathbf{K'}$ valleys with opposite directions which
results in valley dependent optical selection rules. The $\mathbf{K}$ and $\mathbf{K'}$ valleys can be selectively excited by
$\sigma^\pm$ circularly polarized photons, respectively\,\cite{yao2008valley}. In addition, the strong spin-orbit interaction in TMDs lifts the
degeneracy of spin states particularly in the valence band giving rise to well separated A and B-excitonic transitions in
absorption or reflectance spectra\,\cite{PhysRevLett.105.136805, he2014tightly,dhakal2014confocal,he2013experimental}. The
valence band splitting in TMDs is in the range of a few-hundred meV\,\cite{rev2,rev5}. In the conduction band this splitting is
much smaller and is in the range of few to tens meV\,\cite{rev5,echeverry2016splitting}. Spin-orbit interaction in combination
with the broken time reversal symmetry leads to the coupling of spin states and valleys indexes (spin states are reversed in
$\mathbf{K}$ and $\mathbf{K'}$)\,\cite{xiao2012coupled}.

The spin-orbit interaction in the conduction band lifts the degeneracy of bright and dark exciton states of the A and B-excitons. In figure\,\ref{fig:scheme}
the possible spin configurations of the A-exciton in MoX$_2$ and WX$_2$ (X=S, Se) are indicated schematically. Here, and in the
following, we neglect the B-excitons which do not contribute to PL emission due to the large spin-orbit splitting of the valence
band. The schematic is based on both experiments\,\cite{schmidt2016exciton,zhang2015experimental} and theoretical
calculations\,\cite{liu2013three, echeverry2016splitting, rev5} which show that Molybdenum based TMDs have a bright A-exciton
ground state while Tungsten based TMDs have a dark A-exciton ground state due to the reversal of the spin levels in the
conduction band (the situation is reversed for the B-excitons).

Because of valley spin-locking, the large distance (in $\mathbf{k}$ space) between the $\mathbf{K}$ and $\mathbf{K'}$ valleys and
the significant valence band splitting, it is expected that the exciton scattering between valleys will be strongly suppressed and
robust valley polarization can be achieved. Experimentally, a significant degree of circular polarization of photoluminescence
(PL) after circularly polarized excitation has been observed in MoS$_2$, WS$_2$ and WSe$_2$\,\cite{zeng2012valley,
mak2012control, plechinger2015identification, scrace2015magnetoluminescence, wang2014valley, wang2015polarization,
kioseoglou2012valley}. In addition, coherent emission from $\mathbf{K}$ and $\mathbf{K'}$ valleys was observed in WSe$_2$
\,\cite{jones2013optical}. These results demonstrate that valley polarization persists for a sufficiently long time to be
observed in the polarization resolved PL. However, there are some inconsistencies in this picture.

\begin{figure}
\centering
\includegraphics[width=0.8\linewidth]{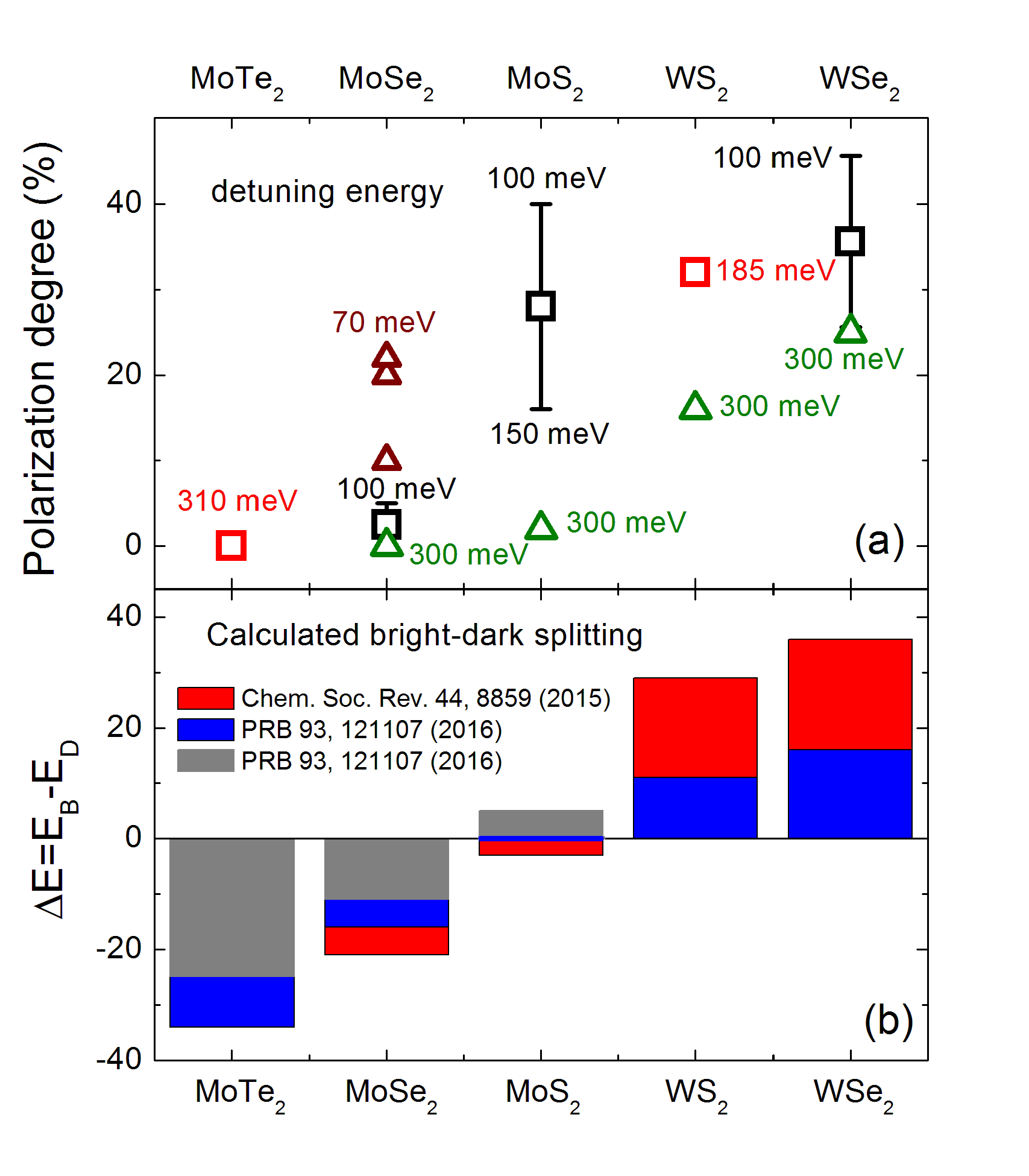}
\caption{(a) Reported literature values of PL circular polarization degree for
TMDs\,\cite{Aivazian2015,mak2012control,jones2013optical,scrace2015magnetoluminescence,wang2014valley,wang2015polarization,zhu2014anomalously,
plechinger2015identification,kioseoglou2012valley,zeng2012valley,lagarde2014carrier,kioseoglou2016optical,arora2016}. The open
squares are average values for detuning range of $100-150$\,meV and error bars represent maximum and minimum of reported values.
Points with different value of detuning are labeled accordingly. (b) Literature data for the calculated difference between bright
and dark exciton state energy\,\cite{rev5,echeverry2016splitting}.} \label{fig:bars}
\end{figure}

Despite the very similar band structure of WX$_2$ and MoX$_2$ monolayers, it is much easier to achieve polarized emission in
Tungsten based TMDs (see for example\,\cite{wang2015polarization}). In particular, obtaining polarized PL emission in MoSe$_2$
proved to be extremely difficult, requiring almost resonant excitation (detuning $\sim 70$\,meV) and gave only 20\% circular
polarization\,\cite{kioseoglou2016optical}. Another puzzling aspect is that in WX$_2$ monolayers the PL polarization has only a
weak dependence\,\cite{jones2013optical, wang2015polarization, zhu2014anomalously} on the excitation photon energy while in
MoX$_2$ the dependence is more pronounced\,\cite{wang2015polarization, kioseoglou2012valley, kioseoglou2016optical, mak2012control}.
These observations are summarized in figure \ref{fig:bars}(a), where the reported values of the degree of circular polarization
for various detunings of the excitation are shown. It is clearly visible that for large value of detuning the PL polarization is
achieved only for WX$_2$. For detuning in the range of $100-150$\,meV only MoS$_2$ exhibits significant degree of circular
polarization in case of Molybdenum based materials.

Here we show that the achievable  degree of circular polarization is directly linked with the alignment of bright and dark
exciton states in Mo and W based monolayer TMDs. We propose a simple rate equation model that demonstrates that the different
alignment of excitonic states can enhance the degree of PL circular polarization which is limited by the inter-valley scattering.
Strong support for this model is provided by a detailed excitation energy dependent investigation of the degree of circular
polarization of the PL emission in MoSe$_2$ for detunings as small as 20\,meV.

\section{Model}

Despite the large valence band splitting and spin-valley locking, ultrafast optical spectroscopy experiments show that the
inter-valley scattering is very effective in TMDs and the valley depolarization occurs in the range of few to several picoseconds
\,\cite{lagarde2014carrier, mai2013many, wang2013valley, zhu2014exciton, dal2015ultrafast}. There are different propositions in
literature explaining exciton scattering between valleys.

One of the proposed mechanisms involves the thermalization of photo-created electron-hole pairs via the emission of two
longitudinal acoustic (LA) phonons\,\cite{kioseoglou2012valley, kioseoglou2016optical,cao2012valley}. Provided that exciton
kinetic energy is above the energy of two LA phonons at the $\mathbf{K}$ point, the exciton can be scattered between the
$\mathbf{K}$ and $\mathbf{K'}$ valleys. Two phonon emission is required for the simultaneous scattering of the electron and hole.
The LA phonon model is particularly appealing as it provides a relatively straightforward way to describe the degree of PL
circular polarization as a function of the excitation energy. The larger LA phonon energy in MoS$_2$
(30\,meV)\,\cite{molina2011phonons} compared to MoSe$_2$ (19\,meV)\,\cite{horzum2013phonon} is consistent with the significant
degree of circular polarization of the PL emission observed in MoS$_2$ for larger detuning of the excitation than in MoSe$_2$.
However, the LA phonon model is unable to explain the different degrees of polarization observed in MoX$_2$ and WX$_2$. The LA
phonon energy of WS$_2$ and WSe$_2$ at the $\mathbf{K}$ point are respectively 24\,meV\,\cite{molina2011phonons} and
18\,meV\,\cite{terrones2014new}. Therefore, we would expect that the observation of polarized PL emission in WX$_2$ should be
extremely difficult (as in MoSe$_2$), while experimentally polarized emission is easily achieved.

An alternative explanation for the valley depolarization is related to electron-hole exchange interaction\,\cite{yu2014valley,
glazov2014exciton}. This mechanism can be considered as the simultaneous virtual recombination of a bright exciton in the
$\mathbf{K}$ valley and the creation of a bright exciton in the $\mathbf{K'}$ valley\,\cite{yu2014valley}. Theoretically, the effectiveness of this
mechanism depends on the exciton center of mass momentum\,\cite{yu2014valley}. The scattering is faster when the exciton has
large kinetic energy, which in principle explains the dependence of the degree of circular polarization of the PL emission on the
energy of the excitation. The differences in the degree of circular polarization in different materials can be attributed to
different material constants. However, in figure\,\ref{fig:bars} there is a clear correlation between the size of the bright
and dark exciton splitting and the obtained degree of polarization. This suggests that a dark exciton ground state favors
polarized emission, even for non resonant excitation.

In the simplest approach, the kinetics of valley polarization can be described by two coupled rate equations
\,\cite{kioseoglou2012valley}:
\begin{equation}
\left\{ \begin{array}{l}
\frac{\rmd N}{\rmd t}=-\frac{N}{\tau_{sk}}-\frac{N}{\tau_r}+\frac{N'}{\tau_{sk}}+g \nonumber\\
 \\
\frac{\rmd N'}{\rmd t}=-\frac{N'}{\tau_{sk}}-\frac{N'}{\tau_r}+\frac{N}{\tau_{sk}}+g' ,\nonumber
\end{array}\right.
\end{equation}
where $N$ and $N'$ are the exciton populations in the $\mathbf{K}$ and $\mathbf{K'}$ valleys, $\tau_{sk}$ is the intervalley
(exchange) scattering time, $\tau_{r}$ is the recombination time of exciton (including both radiative and nonradiative
recombination channels) and $g$ and $g'$ are generation rates of excitons in the $\mathbf{K}$ and $\mathbf{K'}$ valley. In this
model the time dependent degree of the PL circular polarization $P(t)$ is given by
\begin{equation}
P(t)=\frac{N(t)-N(t)'}{N(t)+N(t)'}\nonumber
\end{equation}
For cw excitation ($g=$const or $g'=$const) after some time the population of excitons in the $\mathbf{K}$ and $\mathbf{K'}$
valleys, and thus the degree of circular polarization, are constant. In the following we consider only steady state conditions
and we will not discuss the time dependence of $P$.

Including the effect of exciton scattering between bright and dark states requires four coupled rate equations. It is important
to note that the exchange interactions can scatter excitons only between \emph{bright} exciton states in the $\mathbf{K}$ and
$\mathbf{K'}$ valleys\,\cite{yu2014valley}. We assume, as before, that the inter-valley bright exciton scattering is described by
the time constant $\tau_{sk}$. A term is added to describe the intra-valley scattering between bright and dark states, which has
been observed experimentally\,\cite{zhang2015experimental, schmidt2016exciton}, notably in the temperature dependence of PL
intensity which depends on the alignment of bright and dark exciton states. The PL in MoS$_2$ and MoSe$_2$ is quenched with
increasing temperature while in WSe$_2$ the PL is initially enhanced with increasing temperature. This observation shows that the
alignment of the bright and dark exciton states has significant impact on the exciton recombination kinetics. Moreover, the
transient absorption spectroscopy experiments demonstrate that carriers experience very effective intra-valley scattering with
simultaneous spin flip process\,\cite{mai2014exciton, sie2015intervalley} which can also lead to the population of dark excitons
states. In the WX$_2$ TMDs the scattering rate from dark to bright states is described by
$\frac{1}{\tau_{bd}}\exp\big(-\frac{\Delta E}{kT}\big)$  while in MoX$_2$ TMDs the scattering rate is equal to
$\frac{1}{\tau_{bd}}$. Scattering from bright to dark states are described by $\frac{1}{\tau_{bd}}$ and
$\frac{1}{\tau_{bd}}\exp\big(-\frac{\Delta E}{kT}\big)$ for WX$_2$ and MoX$_2$, respectively. The Boltzmann factor reflects the
experimentally observed energy barrier $\Delta E$ (the absolute value of bright-dark exciton state splitting) for scattering
between bright and dark states\,\cite{zhang2015experimental}. This factor is responsible for the very different exciton
recombination kinetics depending upon the nature of the ground state. Since in WX$_2$ the transition from bright to dark state is
downwards it is effective while in MoX$_2$ this transition is damped due to the energy barrier. The system of rate equations for
WX$_2$ materials can be written as follows,
\begin{equation}
\left\{ \begin{array}{l}
\frac{\rmd N_b}{\rmd t}=-\frac{N_b}{\tau_{sk}}-\frac{N_b}{\tau_{r}}-\frac{N_b}{\tau_{bd}}+\frac{N'_b}{\tau_{sk}}+\frac{N_d}{\tau_{bd}}\exp\big(-
\frac{\Delta E}{kT}\big)+g\\
 \\
\frac{\rmd N'_b}{\rmd t}=-\frac{N'_b}{\tau_{sk}}-\frac{N'_b}{\tau_{r}}-\frac{N'_b}{\tau_{bd}}+\frac{N_b}{\tau_{sk}}+\frac{N'_d}{\tau_{bd}}\exp\big(-
\frac{\Delta E}{kT}\big)+g' \\
\\
\frac{\rmd  N_d}{\rmd t}=-\frac{N_d}{\tau_{r}}-\frac{N_d}{\tau_{bd}}\exp\big(-\frac{\Delta E}{kT}\big)+\frac{N_b}{\tau_{bd}}\\
\\
\frac{\rmd N'_d}{\rmd t}=-\frac{N'_d}{\tau_{r}}-\frac{N'_d}{\tau_{bd}}\exp\big(-\frac{\Delta E}{kT}\big)+\frac{N'_b}{\tau_{bd}}
\end{array}\right.
\label{WX_equation}
\end{equation}
and for MoX$_2$ it has the following form:
\begin{equation}
\left\{ \begin{array}{l}
\frac{\rmd N_b}{\rmd t}=-\frac{N_b}{\tau_{sk}}-\frac{N_b}{\tau_{r}}-\frac{N_b}{\tau_{bd}}\exp\big(-
\frac{\Delta E}{kT}\big)+\frac{N'_b}{\tau_{sk}}+\frac{N_d}{\tau_{bd}}+g\\
 \\
\frac{\rmd N'_b}{\rmd t}=-\frac{N'_b}{\tau_{sk}}-\frac{N'_b}{\tau_{r}}-\frac{N'_b}{\tau_{bd}}\exp\big(-
\frac{\Delta E}{kT}\big)+\frac{N_b}{\tau_{sk}}+\frac{N'_d}{\tau_{bd}}+g' \\
\\
\frac{\rmd N_d}{\rmd t}=-\frac{N_d}{\tau_{r}}-\frac{N_d}{\tau_{bd}}+\frac{N_b}{\tau_{bd}}\exp\big(-\frac{\Delta E}{kT}\big)\\
\\
\frac{\rmd N'_d}{\rmd t}=-\frac{N'_d}{\tau_{r}}-\frac{N'_d}{\tau_{bd}}+\frac{N'_b}{\tau_{bd}}\exp\big(-\frac{\Delta E}{kT}\big),
\end{array}\right.
\label{MX_equation}
\end{equation}

The inter and intra valley scattering and recombination paths involved are summarized in figure\,\ref{fig:scheme}. In writing the
rate equations we have assumed identical recombination times for the dark and bright excitons which is reasonable in TMDs due to
the strong non radiative recombination\,\cite{Wang2015,Wang2015a} which dominates over the radiative recombination of bright
excitons. Under steady state conditions ($dN/dt=0$) if we consider only the bright-dark exciton scattering in the rate equations
we obtain $N_b/N_d = \exp(-\Delta E/kT)$ for WX$_2$ and $N_d/N_b = \exp(-\Delta E/kT)$ for MoX$_2$ which is simply the expected
Boltzmann distribution. Thus, when the $\tau_{bd}$ is similar or shorter than $\tau_r$ (which is the case in TMDs \emph{i.e.}
$\tau_{bd}\ll 1ps$\,\cite{song2013transport, mai2014exciton, sie2015intervalley} and $\tau_r$ is in the range of picoseconds to
hundreds of ps\,\cite{zhang2015experimental, schaibley2015population, yan2014photoluminescence, Wang2015, Wang2015a, korn2011low,
robert2016exciton, amani2015near, schmidt2016exciton}) and the dark exciton is the ground state (in WX$_2$ TMDs) it provides an
important reservoir for valley polarization which tries to maintain the bright exciton population in the same valley (\emph{i.e.}
tries to maintain a Boltzmann distribution). This reservoir is robust since there is no exchange inter-valley scattering channel
for dark excitons\,\cite{yu2014valley}.

The observed degree of circular polarization in PL is related to the occupation of exciton bright states in $\mathbf{K}$ and
$\mathbf{K'}$ valleys,
\begin{equation}
P=\frac{N_b-N'_b}{N_b+N'_b}. \label{eq:polar}\nonumber
\end{equation}
To obtain the polarization we solve the rate equations assuming that excitons are generated only in $\mathbf{K}$ valley ($g'=0$),
corresponding to $\sigma^+$ excitation. The generation rate is assumed to be constant (cw excitation) and the polarization at
steady state conditions ($dN/dt = 0$) is calculated. For simplicity all time constants for different processes are expressed in
units of the exciton recombination time $\tau_{r}$ and the temperature is assumed to be $T=10$\,K (most of the PL literature data
are taken close to this temperature).

\begin{figure}
\centering
\includegraphics[width=1\linewidth]{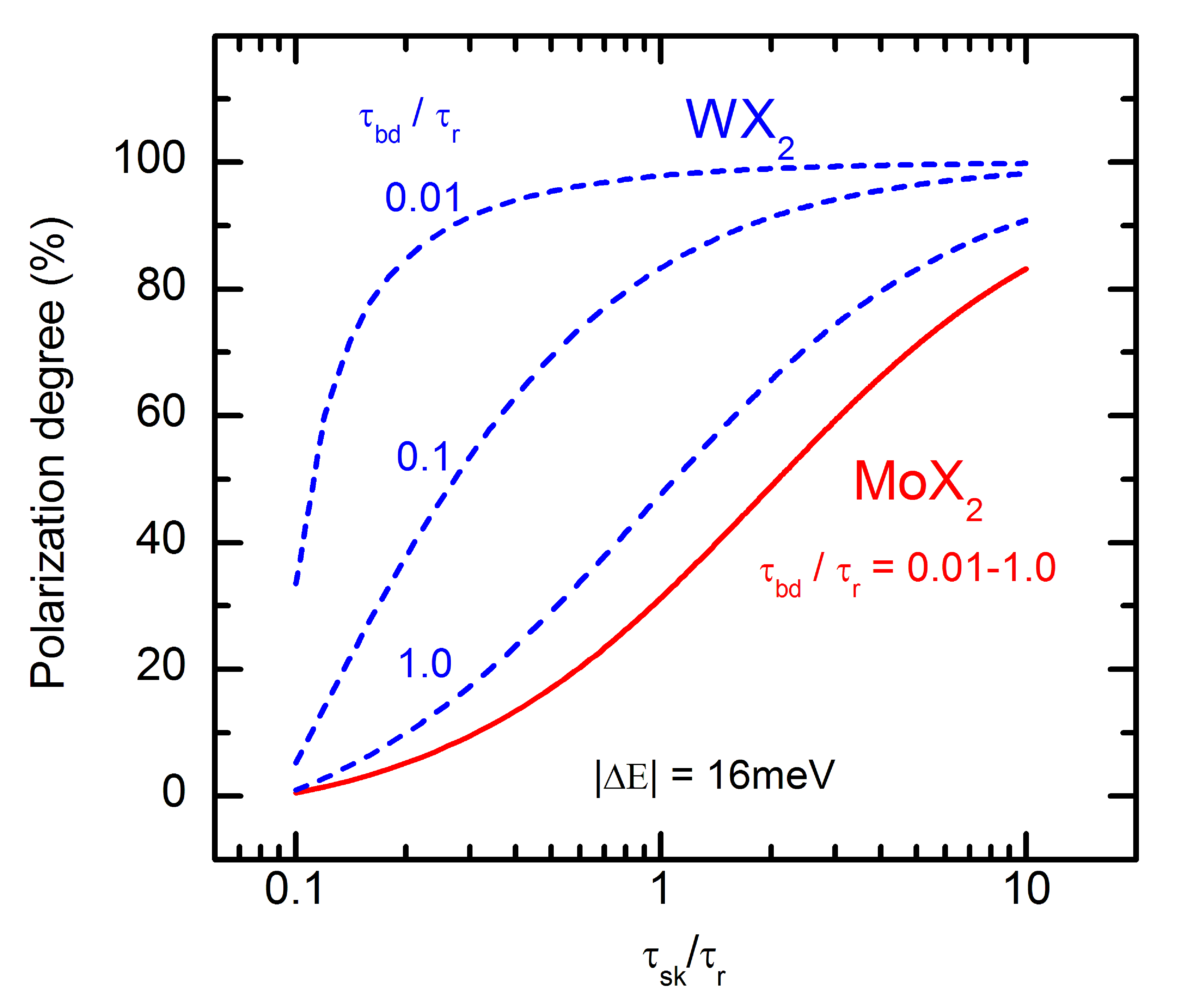}
\caption{Degree of circular polarization in PL as a function of the
inter-valley scattering time calculated using the rate equations for
WX$_2$ (broken lines) and MoX$_2$ (solid lines) for three different
values of the bright-dark exciton scattering time $\tau_{bd}$.}
\label{fig:model1}
\end{figure}

Figure \ref{fig:model1} shows the calculated degree of polarization as a function of the inter-valley scattering time $\tau_{sk}$
for WX$_2$ and MoX$_2$ TMDs, calculated for three different values of $\tau_{bd}$ assuming a typical bright-dark splitting of
$\Delta E = 16$\,meV. It is clearly seen that with decreasing $\tau_{sk}$ the degree of polarization decreases rapidly for both
material systems as expected. However, the degree of polarization for TMDs with a dark exciton ground state (WX$_2$) is always
higher than for TMDs with a bright exciton ground state (MoX$_2$). This has to be related to the different efficiency of the
bright-dark exciton scattering channel, since the other terms in the rate equations are identical. Excitons are created only in
the bright state and then scattered either to the other valley or to a dark exciton state in the same valley. Since the second
process is highly efficient in WX$_2$ (dark exciton ground state), it competes with inter-valley scattering. Thus, fewer excitons
are scattered to the opposite valley, and the dark excitons constitute an important reservoir which tries to maintain $N_b/N_d$
close to the expected Boltzmann distribution. This increases the degree of polarization of the PL emission (and also the total
valley polarization). For MoX$_2$, the bright-dark exciton scattering is strongly damped and the inter-valley scattering
dominates, which depletes the population of the optically excited valley, decreasing the degree of polarization of the PL
emission. For MoX$_2$ TMDs, changes in the scattering time between bright and dark states do not affect the polarization degree
(all lines overlap), the kinetics is dominated by inter-valley scattering at low temperature. On the other hand, the value of
$\tau_{bd}$ significantly affects the polarization degree if the ground state is dark (WX$_2$). Fast bright-dark scattering
(small value of $\tau_{bd}$) greatly enhances the degree of polarization especially if the inter-valley bright exciton scattering
is efficient ($\tau_{sk}/\tau_r <1$). It is also notable that when the bright-dark scattering is efficient the polarization
dependence exhibits a plateau for some range of $\tau_{sk}$ values. This is in agreement with experimental observation where for
WX$_2$ material system the degree of circular polarization is not so sensitive to the excitation energy (according to ref.
\,\cite{yu2014valley} while the inter-valley scattering rate increases with exciton kinetic energy).

\begin{figure}
\centering
\includegraphics[width=1\linewidth]{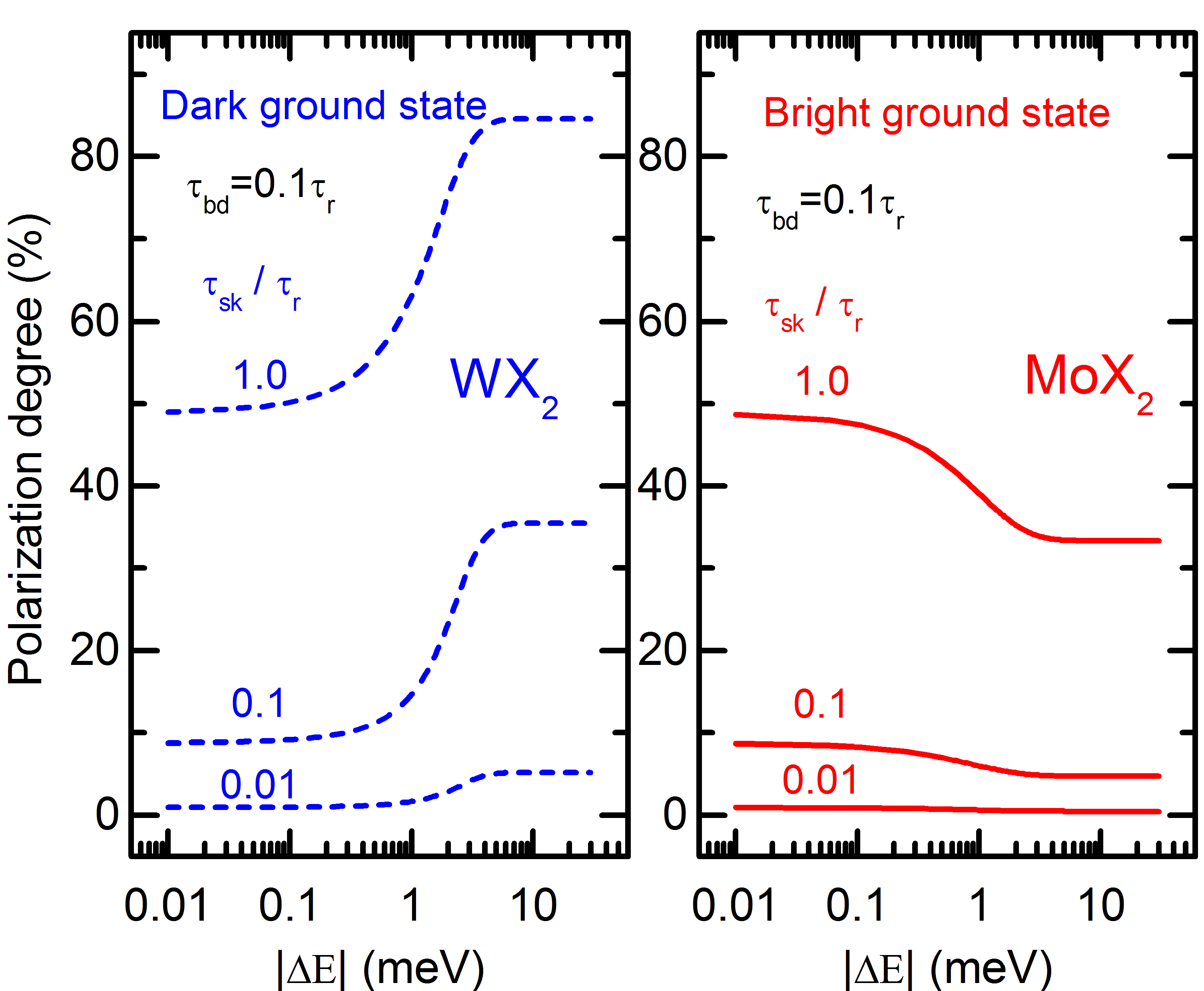}
\caption{Dependence of PL circular polarization degree as function
of absolute value of bright-dark exciton spin splitting calculated
according to rate equations   \ref{WX_equation} (WX$_2$) and
\ref{MX_equation} (MoX$_2$) for different values of inter-valley
scattering time.} \label{fig:model2}
\end{figure}

Figure\,\ref{fig:model2} shows the dependence of the calculated PL polarization degree as a function of the absolute value of
bright-dark exciton splitting for WX$_2$ and MoX$_2$. While the dark and bright states are degenerate ($|\Delta E|\ll kT$) there
is no difference in the PL polarization for either system. When the splitting between bright and dark states becomes non
negligible, the PL circular polarization degree is significantly higher for the WX$_2$ TMDs. With increasing bright-dark
splitting the reservoir of dark states is increasingly populated and the degree of circular polarization in the PL emission
increases. In contrast, in the MoX$_2$ TMDs, increasing the bright-dark splitting reduces the population of the dark exciton
reservoir and the degree of polarization of the PL emission actually decreases. This observation very well explains the observed dependence of PL circular polarization from the TMDs alloys such us Mo$_{1-x}$W$_x$Se$_2$ where the rapid increase of PL circular polarization is observed for increasing Tungsten content\,\cite{wang2015spin} and related to this realignment of bright and dark exciton states.

\begin{figure*}
\centering
\includegraphics[width=0.9\linewidth]{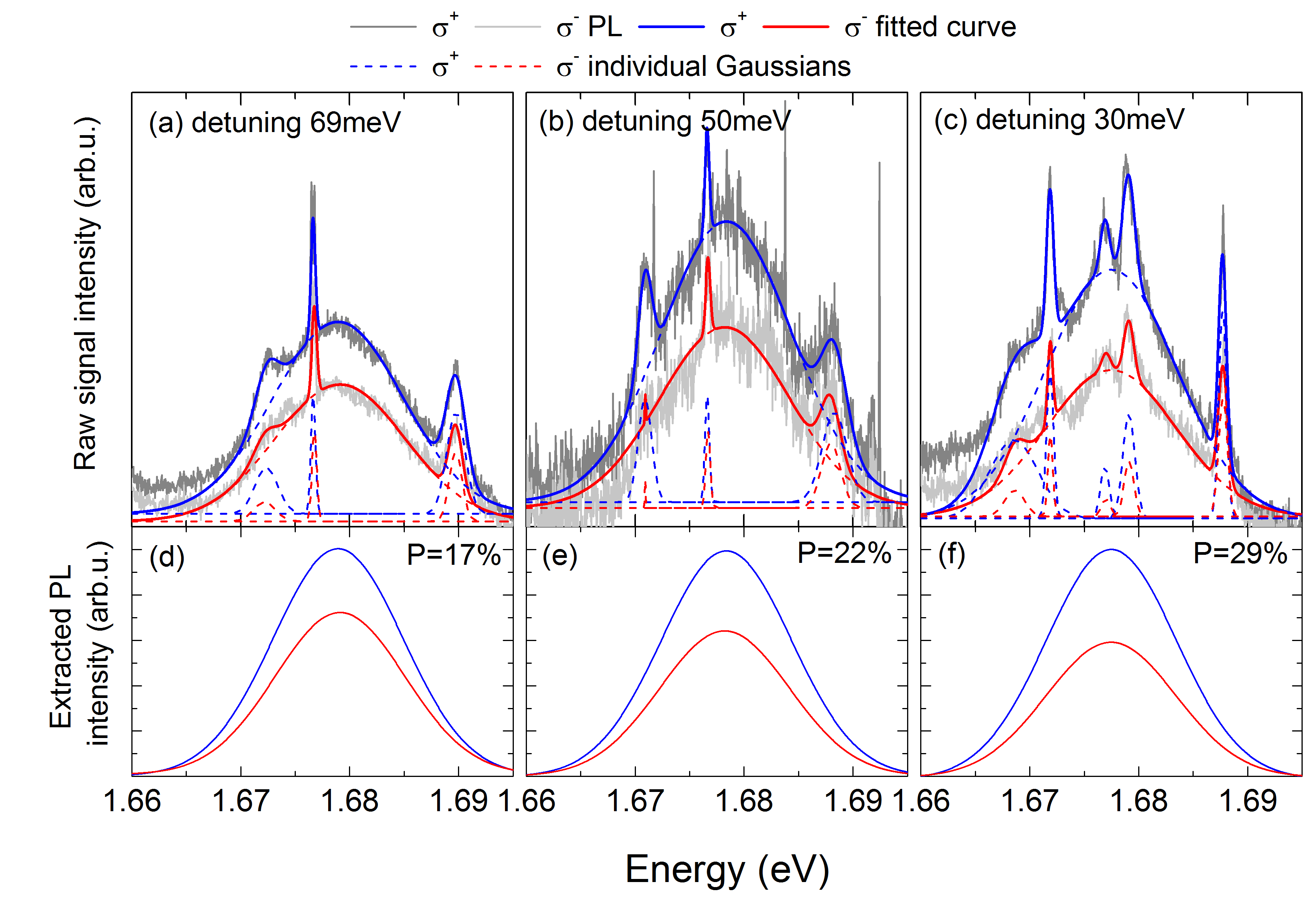}
\caption{(a)-(c)Representative PL spectra obtained for $\sigma^+$ (dark gray) and $\sigma^-$ (light gray) polarization together
with fitted curves (blue and red solid lines), the individual Gaussian components are plotted by dashed lines. (d)-(f) Extracted
component related to PL of the A-exciton, blue (red) color indicates  $\sigma^+$ ( $\sigma^-$) detection polarization. The degree
of circular polarization ($P$) calculated from the amplitudes of the $\sigma^\pm$ PL is indicated on each figure.} \label{fig:PL1}
\end{figure*}

\section{Degree of circular polarization in Molybdenum diselenide}

A crucial test of the rate equation model and the pertinence of the assumption that the electron-hole exchange is the dominant
inter valley scattering mechanism is provided by an investigation of the degree of polarization of the PL emission in MoSe$_2$
which has a bright exciton ground state. For resonant excitation the two LA phonon model predicts 100\% polarization while the
rate equation model predicts significantly less, depending upon the bright-dark exciton scattering rate. For the measurements we
have used a large area CVD grown MoSe$_2$ monolayer encapsulated\,\cite{Wang2016} with monolayer MoS$_2$ to improve the optical
quality\,\cite{Surrente2016}.

For polarization resolved PL studies the sample was placed in a helium cryostat. All measurements were performed at $T \simeq
4.5$\,K and the sample was excited using a tunable CW Ti:sapphire laser. The wavelength of excitation was tuned from 710\,nm to
730\,nm (a much lower energy than the A-exciton transition in MoS$_2$ encapsulating layers) and the excitation power was $\sim
800\,\mu$W. The excitation beam was focused on the sample via a $20\times$ microscope objective (0.55 numerical aperture). The PL
signal was collected through the same objective and dispersed by a triple grating monochromator equipped with 1800\,grooves/mm
grating and detected using a nitrogen cooled CCD detector. The excitation beam was circularly polarized using a linear polarizer
combined with a Babinet-Soleil compensator. Using a zero-order quarter-wave plate and a subtractive mode linear polarizer, the PL
signal was detected in $\sigma^\pm$ polarizations.

\begin{figure}
\centering
\includegraphics[width=1\linewidth]{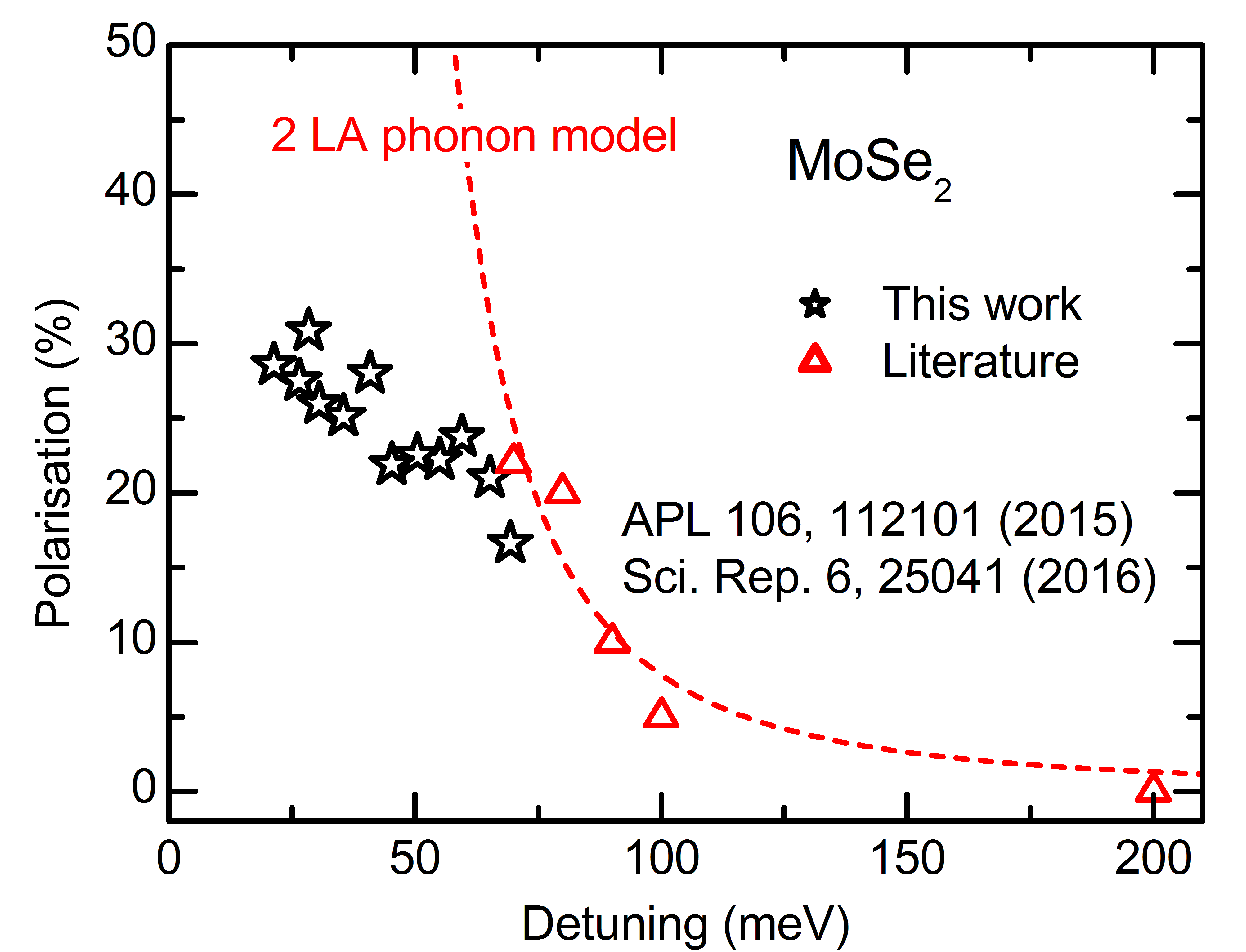}
\caption{Measured degree of circular polarization of PL emission from MoSe$_2$ as a function of detuning between excitation and
exciton transition energy. Values at large detuning (triangles) are taken from the
literature\,\cite{wang2015polarization,kioseoglou2016optical}. The dashed line is calculated dependence for two LA phonon model
using $\hbar\omega_{2LA}=38$\,meV and $\alpha=0.1$.} \label{fig:PL2}
\end{figure}

Representative PL spectra of the A-exciton emission detected in $\sigma^+$ (dark grey line) and $\sigma^-$ (light grey line)
polarization are shown in figure\,\ref{fig:PL1}(a-c). Due to the small difference between the excitation and emission energy, the
PL spectra exhibit, in addition to the A-exciton emission, a number of Raman peaks which progressively strengthen as resonant
excitation is approached. To correctly analyze the degree of circular polarization of the A-exciton PL emission the Raman
features have to be removed. To achieve this we first measure the PL far from resonances using a frequency doubled Nd:YAG laser
emitting at 532\,nm. The PL spectra are then fitted using a Gaussian to extract the broadening and energy of the A-exciton
recombination. These parameters are then fixed when fitting spectra measured close to resonant excitation using a number of
Gaussian functions (depending on the number of Raman features observed). Fixing the energy and broadening of the A-exciton
transition allows the amplitude to be reliably extracted. The results of the fitting procedure are shown by the blue ($\sigma^+$
polarization) and red ($\sigma^-$ polarization) curves in figure \ref{fig:PL1} (a-b) together with individual Gaussians (dashed
lines). We have good agreement between fitted curves and experimental data. The extracted Gaussian peaks (normalized by the
maximum value of $\sigma^+$ PL signal) corresponding to the A exciton transition are presented in figure \ref{fig:PL1}(d-f). The
increase of the $\sigma^+$ PL intensity relatively to $\sigma^-$ is clearly visible with decreasing detuning \emph{i.e.} the
polarization degree increases. Based on the extracted PL intensities in both polarizations we determine the degree of PL circular
polarization for all excitation wavelengths. The dependence of PL circular polarization versus detuning energy is presented in
figure \ref{fig:PL2}. Black stars are our data and red triangles represent values reported for MoSe$_2$ in the
literature\,\cite{kioseoglou2016optical, wang2015polarization}. While the literature data presents a rapid increase in the
polarization for detuning energies below $\simeq 100$meV, our data reveals that this behavior rapidly saturates with an
extrapolated (maximum) polarization for resonant excitation of $\simeq35$\%.

Finally, we compare the measured polarization with the predictions of the two LA phonon model\,\cite{kioseoglou2016optical},
\begin{equation}
P=\left(1+\frac{\alpha \hbar\omega_{2LA} }{\exp(\hbar\omega_{2LA}/( E_{det} - \hbar\omega_{2LA}))-1}\right)^{-1},\nonumber
\end{equation}
where $\hbar\omega_{2LA}$ is twice the LA phonon energy, $E_{det}$ is the detuning and $\alpha \sim 1$ is a scaling constant. The
dashed line in figure\,\ref{fig:PL2} is calculated using $\hbar\omega_{2LA}=38$\,meV and $\alpha=0.1$ which provides the best fit
to the literature data. The two LA phonon model diverges asymptotically to 100\% polarization as the detuning energy approaches
$\hbar\omega_{2LA}$ from above while the experimental data clearly saturates around the 35\% for detunings below
$\hbar\omega_{2LA}$. This strongly suggests that the changes of the PL circular polarization degree with excitation wavelength
are explained by the dependence of the exchange interaction on the exciton kinetic energy\,\cite{yu2014valley}. It is interesting
to note that in case of MoS$_2$ the PL circular polarization degree can reach almost 100\% for small values of detuning
indicating that inter-valley scattering ($1/\tau_{sk}$) is unable to destroy valley polarization if the excitons have low kinetic
energy. In contrast, our experimental results suggest that 100\% polarization cannot be achieved in MoSe$_2$, a clear indication
that inter-valley scattering remains effective even for excitons with low kinetic energy. These observations are consistent with
the alignment of bright and dark exciton states in the Mo based TMDs. For MoSe$_2$ all calculations predict that the bright
exciton is the ground sate and the dark exciton is several meV above (see figure\,\ref{fig:bars}). As we show such alignment results in attenuation of PL circular polarization. At the same time for MoS$_2$
the exciton splitting is close to zero (see figure\,\ref{fig:bars}) so higher value of $P$ is expected (see figure \ref{fig:model2}) than in case of MoSe$_2$.

\section{Conclusions}

In conclusion, we have developed a simple rate equation model for the excitons kinetics which is able to explain the differences
in the observed degree of circular polarization of the PL emission in different TMDs monolayers. Our work suggests that the dark
exciton states play an important, and previously unsuspected role in determining the degree of polarization of the PL emission.
Inter valley scattering, which is mediated by the electron-hole exchange interaction, is effective only for bright exciton
states. A dark exciton ground state provides a robust reservoir for valley polarization, which tries to maintain a Boltzmann
distribution of the bright exciton states in the same valley via intra valley bright dark exciton scattering mechanism. This
provides a simple explanation for the achievable polarization in TMDs. In WX$_2$ TMDs in which the dark exciton is the ground
state, 20-40\% polarization can be achieved, even with non resonant excitation. The MoX$_2$ TMDs with bright exciton ground
states require almost resonant excitation to achieve the same polarization
degree. The proposed model also explains the difference between different Molybdenum based materials as well as TMDs alloys. The dependence of the degree of polarization on the detuning energy of the excitation in MoSe$_2$ suggests
that the electron-hole exchange interaction dominates over two LA phonon emission mechanism for inter valley scattering in TMDs.

%The optimum case is MoS$_2$ in which the dark
%and bright states are nearly degenerate and close to 100\% polarization of the %PL emission can be achieved with near resonant
%excitation. This is in stark contrast to MoSe$_2$ in which the polarization is %limited to around 35\% even under resonant
%excitation conditions.

\ack This work was partially supported by ANR JCJC project milliPICS, the R\'egion Midi-Pyr\'en\'ees under contract MESR
13053031, BLAPHENE project under IDEX program Emergence, HFML-RU/FOM, a member
of the European Magnetic Field Laboratory (EMFL), STCU project 5809 and the Swiss SNF Sinergia Grant no. 147607.\\\\

%\bibliographystyle{iopart-num}
%\bibliography{BibTmds}

\providecommand{\newblock}{}
\begin{thebibliography}{10}
\expandafter\ifx\csname url\endcsname\relax
  \def\url#1{{\tt #1}}\fi
\expandafter\ifx\csname urlprefix\endcsname\relax\def\urlprefix{URL }\fi
\providecommand{\eprint}[2][]{\url{#2}}
% Bibliography created with iopart-num v2.1
% /biblio/bibtex/contrib/iopart-num

\bibitem{wang2012electronics}
Wang Q~H, Kalantar-Zadeh K, Kis A, Coleman J~N and Strano M~S 2012 {\em Nature
  nanotechnology\/} {\bf 7} 699--712

\bibitem{xu2014spin}
Xu X, Yao W, Xiao D and Heinz T~F 2014 {\em Nature Physics\/} {\bf 10} 343--350

\bibitem{rev2}
Zeng H and Cui X 2015 {\em Chem. Soc. Rev.\/} {\bf 44}(9) 2629--2642

\bibitem{rev1}
Duan X, Wang C, Pan A, Yu R and Duan X 2015 {\em Chem. Soc. Rev.\/} {\bf
  44}(24) 8859--8876

\bibitem{rev5}
Liu G~B, Xiao D, Yao Y, Xu X and Yao W 2015 {\em Chem. Soc. Rev.\/} {\bf 44}(9)
  2643--2663

\bibitem{PhysRevLett.105.136805}
Mak K~F, Lee C, Hone J, Shan J and Heinz T~F 2010 {\em Phys. Rev. Lett.\/} {\bf
  105}(13) 136805

\bibitem{zeng2013optical}
Zeng H, Liu G~B, Dai J, Yan Y, Zhu B, He R, Xie L, Xu S, Chen X, Yao W and Cui
  X 2013 {\em Scientific Reports\/} {\bf 3} 1608

\bibitem{splendiani2010emerging}
Splendiani A, Sun L, Zhang Y, Li T, Kim J, Chim C~Y, Galli G and Wang F 2010
  {\em Nano letters\/} {\bf 10} 1271--1275

\bibitem{lopez2014light}
Lopez-Sanchez O, Alarcon~Llado E, Koman V, Fontcuberta~i Morral A, Radenovic A
  and Kis A 2014 {\em Acs Nano\/} {\bf 8} 3042--3048

\bibitem{jariwala2014emerging}
Jariwala D, Sangwan V~K, Lauhon L~J, Marks T~J and Hersam M~C 2014 {\em ACS
  nano\/} {\bf 8} 1102--1120

\bibitem{ross2014electrically}
Ross J~S, Klement P, Jones A~M, Ghimire N~J, Yan J, Mandrus D, Taniguchi T,
  Watanabe K, Kitamura K, Yao W {\em et~al.\/} 2014 {\em Nature
  nanotechnology\/} {\bf 9} 268--272

\bibitem{zhang2014electrically}
Zhang Y, Oka T, Suzuki R, Ye J and Iwasa Y 2014 {\em Science\/} {\bf 344}
  725--728

\bibitem{mak2016photonics}
Mak K~F and Shan J 2016 {\em Nature Photonics\/} {\bf 10} 216--226

\bibitem{xiao2010berry}
Xiao D, Chang M~C and Niu Q 2010 {\em Reviews of modern physics\/} {\bf 82}
  1959

\bibitem{xiao2012coupled}
Xiao D, Liu G~B, Feng W, Xu X and Yao W 2012 {\em Physical Review Letters\/}
  {\bf 108} 196802

\bibitem{cao2012valley}
Cao T, Wang G, Han W, Ye H, Zhu C, Shi J, Niu Q, Tan P, Wang E, Liu B {\em
  et~al.\/} 2012 {\em Nature communications\/} {\bf 3} 887

\bibitem{liu2013three}
Liu G~B, Shan W~Y, Yao Y, Yao W and Xiao D 2013 {\em Physical Review B\/} {\bf
  88} 085433

\bibitem{yao2008valley}
Yao W, Xiao D and Niu Q 2008 {\em Physical Review B\/} {\bf 77} 235406

\bibitem{he2014tightly}
He K, Kumar N, Zhao L, Wang Z, Mak K~F, Zhao H and Shan J 2014 {\em Physical
  review letters\/} {\bf 113} 026803

\bibitem{dhakal2014confocal}
Dhakal K~P, Duong D~L, Lee J, Nam H, Kim M, Kan M, Lee Y~H and Kim J 2014 {\em
  Nanoscale\/} {\bf 6} 13028--13035

\bibitem{he2013experimental}
He K, Poole C, Mak K~F and Shan J 2013 {\em Nano letters\/} {\bf 13} 2931--2936

\bibitem{echeverry2016splitting}
Echeverry J, Urbaszek B, Amand T, Marie X and Gerber I 2016 {\em Physical
  Review B\/} {\bf 93} 121107

\bibitem{schmidt2016exciton}
Godde T, Schmidt D, Schmutzler J, A\ss{}mann M, Debus J, Withers F, Alexeev
  E~M, Del Pozo-Zamudio O, Skrypka O~V, Novoselov K~S, Bayer M and Tartakovskii
  A~I 2016 {\em Phys. Rev. B\/} {\bf 94}(16) 165301

\bibitem{zhang2015experimental}
Zhang X~X, You Y, Zhao S~Y~F and Heinz T~F 2015 {\em Physical Review Letters\/}
  {\bf 115} 257403

\bibitem{zeng2012valley}
Zeng H, Dai J, Yao W, Xiao D and Cui X 2012 {\em Nature nanotechnology\/} {\bf
  7} 490--493

\bibitem{mak2012control}
Mak K~F, He K, Shan J and Heinz T~F 2012 {\em Nature nanotechnology\/} {\bf 7}
  494--498

\bibitem{plechinger2015identification}
Plechinger G, Nagler P, Kraus J, Paradiso N, Strunk C, Sch{\"u}ller C and Korn
  T 2015 {\em physica status solidi (RRL)-Rapid Research Letters\/} {\bf 9}
  457--461

\bibitem{scrace2015magnetoluminescence}
Scrace T, Tsai Y, Barman B, Schweidenback L, Petrou A, Kioseoglou G, Ozfidan I,
  Korkusinski M and Hawrylak P 2015 {\em Nature nanotechnology\/} {\bf 10}
  603--607

\bibitem{wang2014valley}
Wang G, Bouet L, Lagarde D, Vidal M, Balocchi A, Amand T, Marie X and Urbaszek
  B 2014 {\em Physical Review B\/} {\bf 90} 075413

\bibitem{wang2015polarization}
Wang G, Palleau E, Amand T, Tongay S, Marie X and Urbaszek B 2015 {\em Applied
  Physics Letters\/} {\bf 106} 112101

\bibitem{kioseoglou2012valley}
Kioseoglou G, Hanbicki A, Currie M, Friedman A, Gunlycke D and Jonker B 2012
  {\em Applied Physics Letters\/} {\bf 101} 221907

\bibitem{jones2013optical}
Jones A~M, Yu H, Ghimire N~J, Wu S, Aivazian G, Ross J~S, Zhao B, Yan J,
  Mandrus D~G, Xiao D {\em et~al.\/} 2013 {\em Nature nanotechnology\/} {\bf 8}
  634--638

\bibitem{kioseoglou2016optical}
Kioseoglou G, Hanbicki A~T, Currie M, Friedman A~L and Jonker B~T 2016 {\em
  Scientific Reports\/} {\bf 6} 25041

\bibitem{zhu2014anomalously}
Zhu B, Zeng H, Dai J, Gong Z and Cui X 2014 {\em Proceedings of the National
  Academy of Sciences\/} {\bf 111} 11606--11611

\bibitem{Aivazian2015}
Aivazian G, Gong Z, Jones A~M, Chu R~L, Yan J, Mandrus D~G, Zhang C, Cobden D,
  Yao W and Xu X 2015 {\em Nat Phys\/} {\bf 11} 148--152 ISSN 1745-2473

\bibitem{lagarde2014carrier}
Lagarde D, Bouet L, Marie X, Zhu C, Liu B, Amand T, Tan P and Urbaszek B 2014
  {\em Physical review letters\/} {\bf 112} 047401

\bibitem{arora2016}
Arora A, Schmidt R, Schneider R, Molas M~R, Breslavetz I, Potemski M and
  Bratschitsch R 2016 {\em Nano Lett.\/} {\bf 16} 3624--3629 ISSN 1530-6984

\bibitem{mai2013many}
Mai C, Barrette A, Yu Y, Semenov Y~G, Kim K~W, Cao L and Gundogdu K 2013 {\em
  Nano letters\/} {\bf 14} 202--206

\bibitem{wang2013valley}
Wang Q, Ge S, Li X, Qiu J, Ji Y, Feng J and Sun D 2013 {\em ACS nano\/} {\bf 7}
  11087--11093

\bibitem{zhu2014exciton}
Zhu C, Zhang K, Glazov M, Urbaszek B, Amand T, Ji Z, Liu B and Marie X 2014
  {\em Physical Review B\/} {\bf 90} 161302

\bibitem{dal2015ultrafast}
Dal~Conte S, Bottegoni F, Pogna E, De~Fazio D, Ambrogio S, Bargigia I, D'Andrea
  C, Lombardo A, Bruna M, Ciccacci F {\em et~al.\/} 2015 {\em Physical Review
  B\/} {\bf 92} 235425

\bibitem{molina2011phonons}
Molina-Sanchez A and Wirtz L 2011 {\em Physical Review B\/} {\bf 84} 155413

\bibitem{horzum2013phonon}
Horzum S, Sahin H, Cahangirov S, Cudazzo P, Rubio A, Serin T and Peeters F 2013
  {\em Physical Review B\/} {\bf 87} 125415

\bibitem{terrones2014new}
Terrones H, Del~Corro E, Feng S, Poumirol J, Rhodes D, Smirnov D, Pradhan N,
  Lin Z, Nguyen M, Elias A {\em et~al.\/} 2014 {\em Scientific reports\/} {\bf
  4} 4215

\bibitem{yu2014valley}
Yu T and Wu M 2014 {\em Physical Review B\/} {\bf 89} 205303

\bibitem{glazov2014exciton}
Glazov M, Amand T, Marie X, Lagarde D, Bouet L and Urbaszek B 2014 {\em
  Physical Review B\/} {\bf 89} 201302

\bibitem{mai2014exciton}
Mai C, Semenov Y~G, Barrette A, Yu Y, Jin Z, Cao L, Kim K~W and Gundogdu K 2014
  {\em Physical Review B\/} {\bf 90} 041414

\bibitem{sie2015intervalley}
Sie E~J, Frenzel A~J, Lee Y~H, Kong J and Gedik N 2015 {\em Physical Review
  B\/} {\bf 92} 125417

\bibitem{Wang2015}
Wang H, Zhang C and Rana F 2015 {\em Nano Lett.\/} {\bf 15} 8204--8210 ISSN
  1530-6984

\bibitem{Wang2015a}
Wang H, Zhang C and Rana F 2015 {\em Nano Lett.\/} {\bf 15} 339--345 ISSN
  1530-6984

\bibitem{song2013transport}
Song Y and Dery H 2013 {\em Physical review letters\/} {\bf 111} 026601

\bibitem{schaibley2015population}
Schaibley J~R, Karin T, Yu H, Ross J~S, Rivera P, Jones A~M, Scott M~E, Yan J,
  Mandrus D, Yao W {\em et~al.\/} 2015 {\em Physical review letters\/} {\bf
  114} 137402

\bibitem{yan2014photoluminescence}
Yan T, Qiao X, Liu X, Tan P and Zhang X 2014 {\em Applied Physics Letters\/}
  {\bf 105} 101901

\bibitem{korn2011low}
Korn T, Heydrich S, Hirmer M, Schmutzler J and Sch{\"u}ller C 2011 {\em Applied
  Physics Letters\/} {\bf 99} 102109

\bibitem{robert2016exciton}
Robert C, Lagarde D, Cadiz F, Wang G, Lassagne B, Amand T, Balocchi A, Renucci
  P, Tongay S, Urbaszek B {\em et~al.\/} 2016 {\em Physical Review B\/} {\bf
  93} 205423

\bibitem{amani2015near}
Amani M, Lien D~H, Kiriya D, Xiao J, Azcatl A, Noh J, Madhvapathy S~R, Addou R,
  Santosh K, Dubey M {\em et~al.\/} 2015 {\em Science\/} {\bf 350} 1065--1068

\bibitem{wang2015spin}
Wang G, Robert C, Suslu A, Chen B, Yang S, Alamdari S, Gerber I~C, Amand T,
  Marie X, Tongay S {\em et~al.\/} 2015 {\em Nature communications\/} {\bf 6}
  10110

\bibitem{Wang2016}
Wang K, Huang B, Tian M, Ceballos F, Lin M~W, Mahjouri-Samani M, Boulesbaa A,
  Puretzky A~A, Rouleau C~M, Yoon M, Zhao H, Xiao K, Duscher G and Geohegan D~B
  2016 {\em ACS Nano\/} {\bf 10} 6612--6622 ISSN 1936-0851

\bibitem{Surrente2016}
Surrente A {\em et~al.\/} 2016  The detailed sample preparation and optical
  caracterisation of the encapsulated samples will be the subject of another
  manuscript which is under preparation.

\end{thebibliography}

\providecommand{\newblock}{}

\end{document}